\documentclass[twocolumn,nofootinbib]{revtex4}%

\usepackage{amssymb}
\usepackage{amsfonts}
\usepackage{amsmath}
\usepackage{graphicx,color}
\usepackage{slashed}
\usepackage{young}
\usepackage[vcentermath]{youngtab}
\usepackage[utf8]{inputenc}

\usepackage{tensor}
\usepackage{hyperref}

\makeatletter

\newcommand{\be}{\begin{equation}}
\newcommand{\ee}{\end{equation}}
\newcommand{\bea}{\begin{eqnarray}}
\newcommand{\eea}{\end{eqnarray}}

\newcommand*\xbar[1]{%
  \hbox{%
    \vbox{%
      \hrule height 0.5pt 
      \kern0.3ex
      \hbox{%
        \kern-0.0em
        \ensuremath{#1}%
        \kern-0.0em
      }%
    }%
  }%
} 

\begin{document}

\title{Asymptotic symmetry algebra of Einstein gravity and Lorentz generators}
\author{Oscar Fuentealba${}^{1}$}\email{oscar.fuentealba@ulb.be}
\author{Marc Henneaux${}^{1,2}$}\email{marc.henneaux@ulb.be}
\author{C\'edric Troessaert${}^{1,3}$}\email{cedric.troessaert@hers.be}
\affiliation{${}^1$Universit\'e Libre de Bruxelles and International Solvay Institutes, ULB-Campus Plaine CP231, B-1050 Brussels, Belgium}
\affiliation{${}^2$Coll\`ege de France, 11 place Marcelin Berthelot, 75005 Paris, France}
\affiliation{${}^3$Haute-Ecole Robert Schuman, Rue Fontaine aux M\^{u}res, 13b, B-6800, Belgium}

\begin{abstract}
{The asymptotic symmetry algebra of four-dimensional Einstein gravity in the asymptotically flat context has been shown recently to be the direct sum of the Poincar\'e algebra and of an infinite-dimensional abelian algebra (with central charge) that includes the Bondi-Metzner-Sachs supertranslations.  This result, obtained within the Hamiltonian formalism, yields a supertranslation invariant definition of the Lorentz generators (angular momentum and boosts).  Definitions of Lorentz generators free from the ``supertranslation ambiguities'' have also been proposed recently  at null infinity.  We prove the equivalence of the two approaches for redefining the charges.  
}
\end{abstract}



\maketitle

In a recent paper \cite{Fuentealba:2022xsz}, we showed that the Bondi-Metzner-Sachs symmetry \cite{Bondi:1962px,Sachs:1962wk,Sachs:1962zza} of asymptotically flat spacetimes in four spacetime dimensions could be extended by including a certain class of logarithmic supertranslations.  This was achieved by considering more general asymptotic conditions for the metric $g_{ij}(\mathbf{x})$ and its conjugate momentum $\pi^{ij}(\mathbf{x})$ than those of \cite{Henneaux:2018hdj}, which were shown to provide a non trivial and faithful Hamiltonian realization of the BMS symmetry. These more general boundary conditions schematically take the form
\be
g_{ij}(\mathbf{x}) \longrightarrow_{r \rightarrow \infty}   \Delta_{ij}^{\textrm{log}} + \Delta_{ij}^\textrm{Diff} + g_{ij}^{\textrm{RT}} \label{Eq:BC1}
\ee
\be
\pi^{ij}(\mathbf{x}) \longrightarrow_{r \rightarrow \infty}  \Gamma^{ij}_{\textrm{log}} + \Gamma^{ij}_\textrm{Diff} + \pi^{ij}_{\textrm{RT}} \label{Eq:BC2}
\ee
where (i) $\Delta_{ij}^{\textrm{log}}$ and $\Gamma^{ij}_{\textrm{log}}$ are the changes in the metric and its conjugate momentum due to finite diffeomorphisms of order $\mathcal{O}(\ln r)$ {of the form considered in \cite{Fuentealba:2022xsz}}, (ii) $\Delta_{ij}^\textrm{Diff}$ and $\Gamma^{ij}_\textrm{Diff}$ are the changes in the metric and its conjugate momentum due to finite diffeomorphisms of order $\mathcal{O}(1)$ {of the form considered in \cite{Henneaux:2018hdj}}, and (iii) $g_{ij}^{\textrm{RT}}$ and $\pi^{ij}_{\textrm{RT}}$ obey the boundary conditions of \cite{Regge:1974zd}, namely,
\be g_{ij}^{\textrm{RT}} = \delta_{ij} + \frac{h_{ij}(\mathbf{n})}{r}+o(r^{-1}), \quad \pi^{ij}_{\textrm{RT}} = \frac{p^{ij}(\mathbf{n})}{r^2}+o(r^{-2}), 
\ee
 with $h_{ij}(\mathbf{n})$ even and $p^{ij}(\mathbf{n})$ odd under the antipodal map of the sphere,  $h_{ij}(\mathbf{-n}) = h_{ij}(\mathbf{n})$, $p^{ij}(\mathbf{-n})= - p^{ij}(\mathbf{n})$. Here, $\mathbf{n}$ is the unit normal to the sphere at infinity, $(n^i) = (\sin \theta \cos \varphi, \sin \theta \sin \varphi, \cos \theta)$. The boundary conditions of \cite{Henneaux:2018hdj} have $\Delta_{ij}^{\textrm{log}}=\Gamma^{ij}_{\textrm{log}} =0$ but $\Delta_{ij}^\textrm{Diff}\not= 0$, $\Gamma^{ij}_\textrm{Diff}\not=0$.  Those of \cite{Regge:1974zd} have $\Delta_{ij}^{\textrm{log}}=\Gamma^{ij}_{\textrm{log}} = \Delta_{ij}^\textrm{Diff} =\Gamma^{ij}_\textrm{Diff}=0$.  As it is customary,  $\mathcal O(r^{-n})$ stands for terms with a decay rate equal or faster than $r^{-n}$ in the limit $r\rightarrow \infty$, while by $o(r^{-n})$ we refer to a fall-off strictly faster than $r^{-n}$.
 
 The terms $\Delta_{ij}^{\textrm{log}}$, $\Delta_{ij}^\textrm{Diff}$, $\Gamma^{ij}_{\textrm{log}}$ and $\Gamma^{ij}_\textrm{Diff}$ are respectively of orders  $\mathcal{O}(\frac{\ln r}{r})$, $\mathcal{O}(\frac{1}{r})$, $\mathcal{O}(\frac{\ln r}{r^2})$ and $\mathcal{O}(\frac{1}{r^2})$.  While having the same orders as  $g_{ij}^{\textrm{RT}}-\delta_{ij}$ and $\pi^{ij}_{\textrm{RT}}$, the diffeomorphism terms $\Delta_{ij}^\textrm{Diff}$ and $\Gamma^{ij}_\textrm{Diff}$ have opposite parities, which permits to distinguish them. The explicit expressions for $\Delta_{ij}^{\textrm{log}}$, $\Delta_{ij}^\textrm{Diff}$, $\Gamma^{ij}_{\textrm{log}}$ and $\Gamma^{ij}_\textrm{Diff}$, which can be found in \cite{Fuentealba:2022xsz}, will not be needed here \cite{Footnote0}.

With the enlarged boundary conditions (\ref{Eq:BC1})-(\ref{Eq:BC2}), the asymptotic symmetry is enhanced and contains, besides supertranslations, ``logarithmic supertranslations'' parametrized by one function on the sphere (with no $\ell = 0$ or $\ell = 1$ modes, which act trivially and can be assumed to be absent).  Furthermore, the charges $L^\alpha$ generating the logarithmic supertranslations are canonically conjugate to those of the pure supertranslations $S_\beta$, i.e., $\{ L^\alpha, S_\beta\} = \delta^\alpha_\beta$, with $\alpha$, $\beta$ continuous variables parametrizing the sphere, $(\alpha) \equiv (x^A)$, $ L^\alpha \equiv L(x^A)$ where $x^A$ are the angles \cite{Fuentealba:2022xsz}.  This key feature enables one to disentangle the logarithmic and pure supertranslations {from the Poincar\'e algebra} by a redefinition ``\`a la Darboux'' of the Lorentz generators, in such a way that the redefined Lorentz generators have vanishing brackets with both the logarithmic and pure supertranslations.  These new generators involve quadratic corrections bilinear in the logarithmic and supertranslations charges and are thus not linear combinations of the original generators, evading Lie algebra obstructions pointed out in \cite{Sachs:1962zza}.  That the proper framework for discussing charges and symmetry generators is that of (nonlinear) Poisson manifolds is of course a well-studied chapter of Hamiltonian dynamics.   The charges themselves might in fact provide sometimes ``nonlinear algebras''.  
 Recent nonlinear examples in the BMS context have been given in \cite{Fuentealba:2021xhn,Fuentealba:2021yvo,Fuentealba:2022yqt}, but there exist many earlier ones, see e.g. \cite{Henneaux:1999ib} where a nonlinear redefinition of the charges was also found to be useful.

Now, similar redefinitions of the Lorentz generators were performed recently at null infinity with the same purpose of eliminating the ``supertranslation ambiguity'' 
\cite{Mirbabayi:2016axw,Bousso:2017dny,Javadinezhad:2018urv,Javadinezhad:2022hhl,Compere:2019gft,Compere:2021inq,Compere:2023qoa,Chen:2021szm,Chen:2021kug,Chen:2021zmu,Chen:2022fbu,Footnote1}.   We prove in this letter that  our charges coincide with those given in these works in the {(non-radiative)} past limit of future null infinity (``$u \rightarrow - \infty$''), where the comparison can be made. Since the equivalence of the various definitions of the null-infinity supertranslation-invariant Lorentz charges at the past boundary of future null infinity has been established in the corresponding literature, it is sufficient to compare our approach with any one of them.  For definiteness, we will compare here our approach with the definition given in the appendix E of  \cite{Compere:2023qoa}.

The Hamiltonian charges of the BMS algebra and of the logarithmic supertranslations  involve explicitly the asymptotic fields switched on by performing a logarithmic diffeomorphism, which appear in  $\Delta_{ij}^{\textrm{log}}$ and $\Gamma^{ij}_{\textrm{log}}$ \cite{Fuentealba:2022xsz}.  Such logarithmic diffeomorphisms are not considered in the null infinity analysis, however. For that reason, we shall compare the redefinitions of the charges when $\Delta_{ij}^{\textrm{log}} = \Gamma^{ij}_{\textrm{log}} = 0$ {(which does not imply that the logarithmic charges themselves vanish, see below)}.  This turns out to be rather immediate, because the asymptotic form of the 
 metric and its conjugate momentum reduce then exactly to those studied in \cite{Henneaux:2018hdj} (see also \cite{Henneaux:2019yax}).  
 
With $\Delta_{ij}^{\textrm{log}} = \Gamma^{ij}_{\textrm{log}} = 0$, the asymptotic form of the metric and its conjugate momentum reads explicitly, in  polar coordinates ($r,x^A$),
\begin{align}
g_{rr}  = &1+\frac{1}{r}\xbar h_{rr}+\mathcal O(r^{-2})\,,\quad 
g_{rA}  =\mathcal O(r^{-1})\,,\\
& g_{AB} =r^{2}\xbar g_{AB}+r \xbar h_{AB} + \mathcal O(1)\,,
\end{align}
and
\begin{align}
\pi^{rr} & =\xbar\pi^{rr}+\mathcal O(r^{-1})\,, \quad
\pi^{rA}  =\frac{1}{r}\xbar\pi^{rA}+\mathcal O(r^{-2})\,,\\
&\pi^{AB}  =\frac{1}{r^{2}}\xbar\pi^{AB}+\mathcal O(r^{-3})\,,
\end{align}
where the leading orders of the metric and the momenta are subject to the conditions:
\begin{align}
\xbar h_{rr} & =(\xbar{h}_{rr})^{\text{even}},  \\
\xbar h_{AB}  &=(\xbar h_{AB})^{\text{even}}+2(\xbar D_{A}\xbar D_BU+\xbar g_{AB}U)\,,
\end{align}
and
\begin{align}
\xbar\pi^{rr} & =(\xbar\pi^{rr})^{\text{odd}}-\sqrt{\xbar g}\,\xbar\triangle V\,, \\
\xbar\pi^{rA} & =(\xbar\pi^{rA})^{\text{even}}-\sqrt{\xbar g}\,\xbar D^{A}V\,,\\
\xbar\pi^{AB} & =(\xbar\pi^{AB})^{\text{odd}}+\sqrt{\xbar g}\,(\xbar D^{A}\xbar D^{B}V-\xbar g^{AB}\xbar\triangle V)\,.
\end{align}
Here, $\xbar D_A$ is the covariant derivative on the unit sphere with round metric $\xbar g_{AB}$ and $\xbar\triangle$ its Laplacian. All coefficients in the $1/r$ expansions are functions on the sphere (i.e., functions of the angles $x^A$ only). They are arbitrary except for the parity assignments on $\xbar h_{ij}$ and $\xbar \pi^{ij}$ just written, completed by the conditions that the function $U$ is odd while the function $V$ is even \cite{Henneaux:2018hdj,Henneaux:2019yax}.   Without loss of generality, we can assume that $U$ contains no $\ell=1$ mode and $V$ contains no $\ell=0$ mode, since these drop from the above expressions.  The functions $U$ and $V$ parametrize the terms $\Delta_{ij}^\textrm{Diff}$ and $\Gamma^{ij}_\textrm{Diff}$. 

It was proved in \cite{Henneaux:2018hdj} that shifts of $U$ by an odd function $W^{\text{odd}}$ and of $V$ by an even function $T^{\text{even}}$, which leave these boundary conditions invariant by construction, are precisely the pure BMS supertranslations.  Under the conditons $\Delta_{ij}^{\textrm{log}} = \Gamma^{ij}_{\textrm{log}} = 0$, the corresponding charges reduce to (modulo weakly vanishing bulk terms that we do not write explicitly)
\begin{equation}
Q_{T,W} =\oint d^{2}x\Big[2\sqrt{\xbar g}\,T^{\text{even}}\xbar h_{rr}+2W^{\text{odd}}\Big(\xbar\pi^{rr}-\xbar\pi\Big)\Big]\,, \label{Eq:QBMS0}
\end{equation}
where we have also included the energy (zero mode $\ell = 0$ of $T^{\text{even}}$) and the linear momentum ($\ell = 1$ mode of $W^{\text{odd}}$), which, even though not acting effectively on $U$ and $V$,  have a non trivial action on the other canonical fields.    We use units such that $16 \pi G = 1$. 

Based on the work of \cite{Troessaert:2017jcm}, the Hamiltonian description of the supertranslations in terms of even and odd functions on the sphere has been shown in \cite{Henneaux:2018hdj} to be equivalent to the familiar description at null infinity. Not only do the supertranslation vector fields match, but also their corresponding charges.  

It is easy to see that the vector fields of the homogeneous Lorentz generators are identical.  Since the generators of (true) symmetries are determined in the canonical formalism by the corresponding transformations up to a constant - easily adjusted to zero in our case - , the matching of the Lorentz transformations implies the matching of the Lorentz charges in the Hamiltonian description at spatial infinity with those of the null infinity description provided these are constructed following moment map methods (in the non radiative past limit $u \rightarrow - \infty$ where the charges are integrable and unambiguously defined).  

We will explicitly use the matching procedure that identifies the supertranslations vector fields of the Hamiltonian description on Cauchy hypersurfaces with the supertranslation vector fields at null infinity.  For that reason, we briefly recall it here.
The idea is to go to hyperbolic coordinates following \cite{Ashtekar:1978zz,BeigSchmidt,Compere:2011ve} and  to integrate the asymptotic equations of motion with Cauchy data given on the slice ``hyperbolic time $s =0$''  \cite{Troessaert:2017jcm},\cite{Henneaux:2018hdj}. In hyperbolic coordinates, 
\begin{equation}
	\eta = \sqrt{-t^2 + r^2}, \qquad s = \frac t r.
\end{equation}
the metric asymptotically goes to
\be
ds^2 \rightarrow d \eta^2 + \eta^2  h^{0}_{ab} dx^a dx^b \, ,\qquad (x^a) \equiv (s,x^A)
\ee
where the metric on the unit hyperboloid is
\begin{equation}
	h^{0}_{ab} dx^a dx^b = - \frac{1}{(1-s^2)^2} ds^2 + \frac{1}{(1-s^2)}
	\xbar g_{AB} dx^Adx^B \, .
\end{equation}
The slice $s=0$ is indeed a Cauchy surface {on which the above asymptotic conditions therefore hold}.
To leading order, {the relevant asymptotic analysis becomes linear and the linearized equations are sufficient for our purposes}.   The matching with null infinity of the solutions in hyperbolic coordinates  is then  performed using the method of \cite{Fried1,Friedrich:1999wk,Friedrich:1999ax}.
The Hamiltonian parity conditions on $s=0$ are  equivalent to parity conditions relating the values of the fields at the point $(s, x^A)$ and at its hyperboloid antipodal image $ (-s, -x^A)$ \cite{Compere:2011ve}, in particular, the past of future null infinity ($s=1$) is related with the future of past null infinity ($s=-1$).    

The functions $W^{\text{odd}}$ and $T^{\text{even}}$ that parametrize the supertranslations in the Hamiltonian formalism combine to form an odd function $\omega(s,x^A) = - \omega(-s,-x^A)$ which is a solution of the equation $(\mathcal{D}^2 + 3) \omega = 0$ on the hyperboloid  and matches the null infinity supertranslation parameter $\alpha (x^A) = \lim_{s \rightarrow 1} \sqrt{1-s^2} \omega$ \cite{Troessaert:2017jcm}.  Here, $\mathcal{D}^2$ is the d'Alembertian on the hyperboloid. One has $W^{\text{odd}} = \omega(0,x^A)$ and $T^{\text{even}} = \partial_s \omega(0,x^A)$.  The detailed transformation between $\alpha$ and $W^{\text{odd}}$ and $T^{\text{even}}$ is given in \cite{Henneaux:2018cst}.

For the matching of the supertranslation charges, one notes that the conjugate momenta $\pi^{ij}$ are related to the time derivatives $\partial_s g_{ij}$ in the usual way.  
One then finds that the Hamiltonian variables $\xbar h_{rr}$ and $\xbar\pi^{rr}-\xbar\pi$ combine to form a function $\sigma(s, x^A) = \sigma(-s, -x^A)$ which obeys the same equation as $\omega$ on the hyperboloid, namely, $(\mathcal{D}^2 + 3) \sigma = 0$, but with the opposite parity.  The function $\sigma$ matches the Bondi mass aspect $m(x^A)$ at null infinity \cite{Troessaert:2017jcm}.  The Hamiltonian supertranslation generator \eqref{Eq:QBMS0}
 is thus equal to 
 \be
 Q_{\alpha(T,W)}= 4 \oint d^{2}x \, \sqrt{\xbar g} \,\alpha \, m\,, \label{Eq:S-N}
 \ee
 where $\alpha$ is the supertranslation parameter defined by $T^{\text{even}}$ and $W^{\text{odd}}$.

As we indicated, the terms $\Delta_{ij}^\textrm{Diff}$ and $\Gamma^{ij}_\textrm{Diff}$ in (\ref{Eq:BC1}) and (\ref{Eq:BC2}) are the changes in the metric and its conjugate due to finite diffeomorphisms of order $\mathcal{O}(1)$ respecting the boundary conditions, i.e., the supertranslations.   The odd function $U \equiv U^{\text{odd}}$ and the even function $V \equiv V^{\text{even}}$ provide therefore a parametrization of the orbits of the supertranslation subgroup through given ($g_{ij}^{\textrm{RT}}$,  $\pi^{ij}_{\textrm{RT}}$).   This parametrization is abelian because the nonlinear terms in the diffeomorphism variations of the fields are of highest order in $1/r$.

It follows that one can repeat verbatim for $U^{\text{odd}}$ and $V^{\text{even}}$ the above derivation of the matching of $W^{\text{odd}}$ and $T^{\text{even}}$ with null infinity supertranslations.  The functions $U^{\text{odd}}$ and $V^{\text{even}}$  combine to form an odd function $\psi(s,x^A) = - \psi(-s,-x^A)$ which is a solution of the equation $(\mathcal{D}^2 + 3) \psi = 0$ on the hyperboloid  and matches the null infinity supertranslation parameter $C (x^A) = \lim_{s \rightarrow 1} \sqrt{1-s^2} \psi$ which parametrizes the supertranslation part (``electric parity part'') {$\Delta^\textrm{Diff} C_{AB} = (- 2 \xbar D_A \xbar D_B + \xbar g_{AB} \xbar \triangle ) C$} of the ``Bondi shear'' $C_{AB}$ at null infinity as $u \rightarrow - \infty$.  We recall that the Bondi shear is the leading perturbation term of the angular part of the metric and is explicitly given in terms of $C(x^A)$ by $\lim_{u \rightarrow - \infty} C_{AB}(u,x^C) = (- 2 \xbar D_A \xbar D_B + \xbar g_{AB} \xbar \triangle ) C \, +$  ``magnetic parity part''.   (The magnetic part is often assumed to be zero in this limit but this is not necessary.) One has $U^{\text{odd}} = \psi(0,x^A)$ and $V^{\text{even}} = \partial_s \psi(0,x^A)$.  The detailed transformation between {the parameters $C$ and ($U^{\text{odd}}$, $V^{\text{even}}$) of the supertranslation part of the metric} is the same as the one given in \cite{Henneaux:2018cst} for $\alpha$, $W^{\text{odd}}$ and $T^{\text{even}}$.  The $\ell=0$ and $\ell=1$ modes of $C$ do not contribute to $C_{AB}$ and can be assumed to be zero, in agreement with what we found for $U^{\text{odd}}$ and $V^{\text{even}}$. 

The key observation that enables one to connect the redefinitions of the Lorentz charges at spatial infinity and at null infinity is that the functions $U^{\text{odd}}$ and $V^{\text{even}}$ {\em are also the logarithmic charges} \cite{Fuentealba:2022xsz}. 
One has indeed, for the generators of logarithmic supertranslations (again modulo weakly vanishing bulk terms that we do not write)
\be
Q_{T_{\log},W_{\log}}  =\oint d^{2}x\Big(2\sqrt{\xbar g}\,T_{\log}U^{\text{odd}}-2\sqrt{\xbar g}\,W_{\log}V^{\text{even}}\Big)\,,
\end{equation}
where $T_{\log}$ and $W_{\log}$ are the parameters of the logarithmic supertranslations of \cite{Fuentealba:2022xsz}. 
It is crucial to note that even though the terms $\Delta_{ij}^{\textrm{log}}$ and $\Gamma^{ij}_{\textrm{log}}$ are set to zero in the asymptotic form of the fields, the values of the generators of the logarithmic supertranslations do not vanish.  There is no contradiction. This is similar to the fact that the value of the generators of shifts in $q$ in classical mechanics, namely $p$, does not vanish even if one freezes the coordinate $q$ to $q=0$.  

The above considerations enable one to match immediately the logarithmic supertranslation charges to null infinity quantities: {as we have just explained, these charges, being identical with the supertranslation parameters ($U^{\text{odd}}$, $V^{\text{even}}$)} combine through the hyperboloid function $\psi$ to form the null infinity potential $C(x^A)$ for the electric part of the Bondi shear $C_{AB}$ along the above lines. This completes incidentally the derivation, from the initial data, of the matching conditions between the future of past null infinity and the past of future null infinity.  Besides the matching of the supertranslation parameters and charges recalled above, one has also the correct matching of the $C$ field \cite{Strominger:2013jfa,Strominger:2017zoo} since $C$ coincides with the odd function $\sqrt{1-s^2}\psi$ on the hyperboloid {\cite{Goldstone}} (for a meaningful comparison with the literature, one must take into account the conventions for the orientation of the null generators at past and future null infinities, which depend on the authors).

We can then compare the redefinitions of the Lorentz generators made respectively at null infinity and spatial infinity to achieve supertranslation invariance.  The formulas (9.18) and (9.19) of  \cite{Fuentealba:2022xsz} for the new Lorentz charges $Q_{b,Y}^{\textrm{inv}}$ can be written, 
\be
Q_{b,Y}^{\textrm{inv}} =  Q_{b,Y}^{\textrm{old}} - Q_{T_{b,Y},W_{b,Y}}^{\textrm{corr}}\,,
\ee
where the correcting term $Q_{T_{b,Y},W_{b,Y}}^{\textrm{corr}}$ is the generator of the following field-dependent supertranslation (pure supertranslation and ordinary translation) 
\be
Q_{T_{b,Y},W_{b,Y}}^{\text{corr}} =2 \oint d^{2}x \Big[\sqrt{\xbar g}\,T_{b,Y}\xbar h_{rr}+W_{b,Y}\Big(\xbar\pi^{rr}-\xbar\pi\Big)\Big] \, .
\ee 
(The fields turned on by logarithmic supertranslations (appearing in  $\Delta_{ij}^{\textrm{log}}$ and $\Gamma^{ij}_{\textrm{log}}$) have been set equal to zero in order to perform the comparison.) Here, the parameters 
\be 
(T_{b,Y}, W_{b,Y}) = \delta_M (U,V)
\ee
are the variations of $U$ and $V$ under Lorentz transformations
\be  
W_{b,Y} = \mathcal{L}_Y U - b V, \quad T_{b,Y} =  \mathcal{L}_Y V - (3b + \partial_A \xbar D^A + b \xbar \triangle ) U
\ee
(see formulas (9.16) and (9.17) of  \cite{Fuentealba:2022xsz}).  Even though $V$ has no mode $\ell <2$, $W_{b,Y}$ has a $\ell = 1$ component through $bV$.  This term is necessary to remove the ordinary translations from the brackets of the Lorentz generators with the pure supertranslations, as explained in the general algebraic discussion in section 9.1 of \cite{Fuentealba:2022xsz}.

Applying the translation rules from spatial infinity to null infinity, one first finds that $\delta_M (U,V)$ becomes the Lorentz variation of the corresponding null infinity field $C$, known to be $\delta_M C = Y^A \partial_A C - \frac12 \xbar D_A Y^A C$ where
the $Y^A$'s stand here for all conformal Killing vectors of the $2$-sphere (related to spatial rotations $Y^A_{\text{rot}}$ and boosts $b$ through $Y^A=Y^A_{\text{rot}}-\frac{1}{2}\xbar D^A b$) and where the variation $Y^A \partial_A C - \frac12 \xbar D_A Y^A C$ includes the needed $\ell = 1$ term, i.e.
$
(T_{b,Y}, W_{b,Y})  \rightarrow Y^A \partial_A C - \frac12 \xbar D_A Y^A C.$
Applying now the formula (\ref{Eq:S-N}), one then gets, 
\be
Q_Y^{\textrm{inv}} =  Q_Y^{\textrm{old}} - 4 \oint d^{2}x \, \sqrt{\xbar g} \big(Y^A \partial_A C - \frac12 D_A Y^A C \big) \, m   \label{Eq:NewQ}
\ee
which coincides with the formula (E.6) of \cite{Compere:2023qoa} {(see also formula (12) of \cite{Chen:2022fbu} for spatial rotations)}.  

We can thus conclude that the constructions of supertranslation-invariant Lorentz charges at spatial infinity and null infinity lead to the same redefinitions.  The interest of the Hamiltonian approach, which allows logarithmic supertranslations by including them in the asymptotic form of the fields through the orbit method, is that all terms appearing in the non-linear redefinition of the Lorentz generators have a well-defined symplectic action.  The logarithmic charges $C$ are more than just definite asymptotic field components but are true generators of symmetries, with well-defined Poisson brackets (when supplemented by the appropriate constraint terms in the bulk).  These are the logarithmic supertranslations, i.e., the diffeomorphisms which behave logarithmically at infinity and satisfy the prescribed parity conditions of \cite{Fuentealba:2022xsz}.  Logarithmic supertranslations are on an equal footing as  the pure supertranslations  and one can write equivalently  the correcting term in the redefined Lorentz generators as the generator of logarithmic supertranslations with field-dependent coefficients given by the Lorentz variation of the supertranslation charges (including ordinary translations).

It is gratifying that all roads to the analysis of the asymptotic symmetries lead to the same final redefinitions of the Lorentz generators.   In that context, the importance of including the non trivial diffeomorphism terms  $ \Delta_{ij}^\textrm{Diff}$ and $ \Gamma^{ij}_\textrm{Diff}$ in the asymptotic expressions of the canonical variables in the Hamiltonian formulation should not be underestimated.   These diffeomorphism terms cannot be set equal to zero because they are improper \cite{Benguria:1976in} and can be given a direct physical interpretation.  

Physical observers and detectors are not sitting at $r = \infty$ (where the metric would not be flat in any case in a cosmological context) but at very large distances such that the above developments in $1/r$ make physical  sense.   Suppose that $\Delta_{ij}^\textrm{Diff}= \Gamma^{ij}_\textrm{Diff} = 0 $ initially for a detector that is ``sufficiently far away''.  At some later times, waves hit the detector, after which the gravitational field close to the detector returns to a situation that can again be described in asymptotically flat terms.  Not only will the ADM mass measured by the observer have changed but also $ \Delta_{ij}^\textrm{Diff}$ and $ \Gamma^{ij}_\textrm{Diff}$ will generically not return to zero,  capturing the memory effect  -- well-known indeed to be related to supertranslations as previous studies at null infinity indicate (see e.g. \cite{Strominger:2017zoo}).  For that reason, setting $\Delta_{ij}^\textrm{Diff}= \Gamma^{ij}_\textrm{Diff} = 0$ is not physically acceptable if one wants to describe real observers at very large but finite distances.  An interesting question is whether one could measure a logarithmic memory effect associated with non-vanishing $\Delta_{ij}^{\textrm{log}}$ and $ \Gamma^{ij}_{\textrm{log}}$, which define also improper gauge transformations.

We close this letter with two comments. First, the fact that the asymptotic symmetry has a direct sum  structure ``\`a la Coleman-Mandula'' simplifies the theory of its representations.  The supertranslations simply provide additional independent quantum numbers to the well-known representations of the Poincar\'e algebra.  Second, the discussion can be repeated in the case of electromagnetism, where a similar construction of Lorentz generators free from the angle-dependent gauge transformation ambiguities was performed in \cite{Fuentealba:2023rvf}, following \cite{Henneaux:2018gfi}. What plays the role of the memory field $(U,V) \leftrightarrow C$ is the even phase $\Phi$ that enters the boundary conditions through the $\mathcal O(1)$ angular components of the vector potential {and the odd phase $\xbar \Psi$ that enters its temporal component.  These match the pure phase part $\phi_-$ of $A_A$ at the past of future null infinity \cite{He:2014cra}} as can be seen by applying the rules for matching the gauge transformations derived in \cite{Henneaux:2018gfi}.


\section*{Acknowledgements}
We thank Geoffrey Comp\`ere and Ricardo Troncoso for important discussions. O.F. is grateful to the Coll\`ege de France for kind hospitality while this article was completed. This work was partially supported by  a Marina Solvay Fellowship (O.F.), by FNRS-Belgium (conventions FRFC PDRT.1025.14 and IISN 4.4503.15), as well as by funds from the Solvay Family.


\begin{thebibliography}{99}
\bibitem{Footnote1}
The fact that there is a supertranslation ambiguity in the Lorentz charges is not necessarily a problem in itself, in much the same way as the dependence of the angular momentum on the position about which it is computed (``translation-ambiguity'') is not a problem.  It is just a feature of the BMS algebra as originally defined.  It leads, however, to situations which are more difficult to interpret.  A supertranslation-independent definition is therefore welcome.
\end{thebibliography}
\end{document}